\documentclass[journal,onecolumn]{IEEEtran}
\IEEEoverridecommandlockouts

\usepackage[utf8]{inputenc} 
\usepackage[T1]{fontenc}
\usepackage{verbatim} 
\usepackage{amssymb}
\usepackage{makecell}

\usepackage[font=small]{caption}
\usepackage{cite}
\ifCLASSINFOpdf
  \usepackage[pdftex]{graphicx}
\else
\fi
\usepackage[cmex10]{amsmath}
\usepackage{amssymb}
\usepackage{subfigure}
\usepackage{multirow}
\usepackage{array}
\usepackage[lofdepth,lotdepth]{subfig}
\usepackage[acronym,toc,shortcuts]{glossaries}
\usepackage[usenames, dvipsnames]{color}
\usepackage{url}
\usepackage{csquotes}
\usepackage{amsmath}
\usepackage{float}
\usepackage[acronym,toc,shortcuts]{glossaries}
\makeglossaries
\newacronym{3GPP}{3GPP}{The 3rd Generation Partnership Project }
\newacronym{BC}{BC}{Block Chain - Blok Zincir}
\newacronym{IoT}{IoT}{Internet of Things - Nesnelerin Interneti}
\newacronym{DLT}{DLT}{Dağıtılmış Kayıt Cihazı Teknolojisi}
\newacronym{LP-WAN}{LP-WAN}{Low Power Wide Area Network}
\newacronym{IMSI}{IMSI}{International mobile subscriber identity}
\newacronym{M2M}{M2M}{Machine-to-Machine}
\newacronym{MNO}{MNO}{Mobile Networks Operator}
\newacronym{MŞO}{MŞO}{Mobill Şebeke Operatörü}
\newacronym{QoS}{QoS}{Quality-of-Service}
\newacronym{P2P}{P2P}{Peer-to-Peer}
\newacronym{PKI}{PKI}{Public Key Infrastructure}
\newacronym{OTT}{OTT}{over-the-top}
\newacronym{UE}{UE}{User Equipment}
\newacronym{VoLTE}{VoLTE}{Voice over LTE}

\usepackage[ruled,vlined,noend]{algorithm2e}
\begin{document}
\IEEEpubid{\makebox[\columnwidth]{978-1-7281-7206-4/20/\$31.00 \copyright\ 2020 IEEE\hfill}
\hspace{\columnsep}\makebox[\columnwidth]{}} 



\title{
Fault-Tolerant Strassen-Like Matrix Multiplication  }



\author{Osman B. Güney
and~Suayb S. Arslan
\thanks{O. B. Güney was with the Department
of Electronics Engineering, Sabancı University, Istanbul, Turkey e-mail: osmanberke@sabanciuniv.edu.}
\thanks{S. S. Arslan was with the Department of Computer Engineering, MEF University, Istanbul, Turkey. email: arslans@mef.edu.tr.}
}



\maketitle




\begin{abstract}
In this study, we propose a simple method for fault-tolerant Strassen-like matrix multiplications. The proposed method is based on using two distinct Strassen-like algorithms instead of replicating a given one. We have realized that using two different algorithms, new check relations arise resulting in more local computations. These local computations are found using computer aided search. To improve performance, special parity (extra) sub-matrix multiplications (PSMMs) are generated (two of them) at the expense of increasing communication/computation cost of the system. Our preliminary results demonstrate that the proposed method outperforms a Strassen-like algorithm with two copies and secures a very close performance to three copy version using only 2 PSMMs, reducing the total number of compute nodes by around 24\% i.e., from 21 to 16.
\end{abstract}


\IEEEpeerreviewmaketitle

\IEEEpubidadjcol

\section{Introduction}
Matrix Multiplication (MM) is one of the fundamental operations in many fields of engineering (machine learning, computer graphics and robotics). Due to high dimensions of the matrices, the MM operation may be the dominant factor in the total operational latency. One of the classical ways to reduce computation delay is to distribute the calculations on multiple nodes and allow parallel processing. Other ways include using algorithms which have lower computational complexity \cite{Parallel_Strassen}, \cite{Karstadt}. To get the best of both worlds, algorithms with low complexity must be distributed for parallel processing. 

In 1969, Strassen discovered that MM can be performed recursively with asymptotically less computational complexity than that of the naive way \cite{Strassen1969}. It is shown that computational complexity of MM with 
$2 \times 2$ base case can be reduced from $O(n^{3})$ to $O(n^{\log_27})$ which is due to using 7 multiplications instead of 8. Any algorithm with this recursive structure is called ``Strassen-like algorithm" \cite{Karstadt}, \cite{CENK}.  After Strassen, better algorithms in terms of number of computational steps have been developed to reduce the hidden constant term in the big \textit{O} notation which leads to the reduction in the asymptotical computational complexity \cite{Bini_1979},\cite{WINOGRAD},\cite{LeGall}. For instance, Winograd \cite{WINOGRAD}, has shown that MM can be performed by using 15 additions/subtractions instead of 18 of Strassen's algorithm. Winograd's algorithm is optimal in terms of both multiplications and additions/subtractions for a MM with $ 2 \times 2 $ standard base case, which achieves the lower bounds provided by Probert \cite {Probert}. 

On the other hand in many distributed settings, worker (slave) nodes may become slow or cease operation completely at random times, which are named as \textit{straggler nodes}. To overcome this difficulty and speed up the parallel operation, fault tolerant techniques such as erasure coding are utilized in different contexts \cite{distributed_ml},\cite{d_con}, \cite{Hierarchical}. In that, master node utilizes a subset of the fastest computations by carefully designing \textit{redundant} calculations. These redundant calculations are generated by first partitioning the multiplicant and the multiplier into smaller sub-blocks. Then, based on the coding technique, linear combinations of these sub-blocks are calculated. However unlike standard MM, Strassen-like algorithms use specific sub-blocking and specific subsets to be added/subtracted and multiplied to achieve recursive complexity advantages. So, direct application of previous literature to Strassen-like MM algorithms is not possible. In addition, no previous literature considered the fault tolerant version of Strassen-like MMs, which has been the main motivation of this work. 

To fill this gap, we propose fault tolerant techniques in this study for Strassen-like MMs (specifically for Strassen and Winograd algorithms) through computer aided search, whereby each of the nodes is assumed to perform only one but smaller size MMs. Furthermore, just like original studies, additional parity computations are designed for generating computation redundancy. Finally, performance comparisons are provided with theoretical and Monte Carlo simulations.

\section{Classical Fault-tolerant Coded Matrix Multiplication Methods}

\label{genel_bakis}
The motivation of applying erasure codes to speed up distributed algorithms has found a variety of applications. For instance, an approximate matrix multiplication scheme based on anytime coding is shown to improve the latency performance of the overall computation in \cite{any}. Yet in another study \cite{gradient}, gradient computations are coded to speed-up training phase of machine learning algorithms. Some of the classical erasure codes that can be applied to the matrix multiplication include Maximum Distance Separable (MDS) codes \cite{yu2017polynomial}, Product codes \cite{HD},  Sparse graph codes\cite{SPARSE}, Hierarchichal codes\cite{Hierarchical} and BP-XOR codes \cite{BP-XOR}. 

Suppose that two large matrices are multiplied i.e., {$\textbf{A}^T\textbf{B}$} where $\textbf{A} \in {\mathbb{R}^{m \times n}}$ and $\textbf{B} \in {\mathbb{R}}^{m \times n}$. The $i^{th}$ and $j^{th}$ columns of the matrices \textbf{A} and \textbf{B}, denoted by $a_i \in \mathbb{R}^m$ and $b_j \in \mathbb{R}^m$, respectively, are multiplied (dot product). In other words, computing {$\textbf{A}^T\textbf{B}$} amounts to $n^2$ dot products, where number of workers, namely $N \geq n^2$. Suppose the columns of $\textbf{A}$ and $\textbf{B}$ are divided into $k$ equal size matrices.  This way, the large matrix multiplication problem {$\textbf{A}^T\textbf{B}$} can be viewed as $k$ instances of smaller matrix multiplication problems distributed over independent compute nodes. The latest response from these nodes determine the overall computational time. In MDS-coded computation \cite{yu2017polynomial}, adding redundancy leads to $n > k$ subcomputations and the overall  computational time is determined by the $k^{th}$ fastest worker among the $n$ workers in the same network. Through order statistics, latency analysis can be performed. 

{Product coded scheme proposed in \cite{HD} encodes computating multiple dimensions and achieves an improved coding gain relative to the MDS-coded scheme. Product code is one way of constructing a larger code with small codes as building blocks. For instance, assuming one is given an $(n, k)$ MDS code, a product code can be constructed as follows. We first arrange $k^2$ symbols in a $k$-by-$k$ array. Every row of the array is encoded with a $(n, k)$ MDS code first. Then, each column of the array is encoded with a $(n,k)$ MDS code, resulting in $n \times n$ coded computation block. Finally columns or rows are distributed to $n$ slave nodes for performing the overall computation.  Also, sparse graph codes are proposed \cite{SPARSE} where each worker chooses a random number of input submatrices based on a given degree distribution $\mathcal{P}$; then computes a weighted linear combination where the weights $W_{ij}$ are randomly drawn from a finite set $\mathcal{S}$. When the master node receives a subset of finished tasks such that the coefficient matrix formed by the weights $W_{ij}$ is full rank, it starts to operate a hybrid decoding algorithm between belief propagation decoding and Gaussian elimination to recover the resultant matrix $C$. On the other hand, hierarchical codes decomposes the overall matrix multiplication task into a hierarchy of heterogeneously sized subtasks. The duty to complete each subtask is shared amongst all workers and each subtask is (generally) of a different complexity. The motivation for the hierarchical decomposition is the recognition that more workers will finish the first subtask than the second (or third, forth, etc.) using erasure coding. Earlier subtasks can therefore be designed to be of a higher rate than later subtasks\cite{Hierarchical}.} 

In all of these previous research, matrices are partitioned into smaller subblocks and combinations of these subblocks are selected to be linearly combined according to an erasure coding algorithm. In Strassen-like algorithms, such inherent specific sub-blocking and combining prevents us from using standard approaches particularly while generating redundant computations. To our best knowledge, this is the first attempt towards generating fault-tolerant Strassen-like matrix multiplication methodologies. 

\section{Fault-Tolerant Strassen-Like  Algorithms}
\label{genel_bakis}

\subsection{Background}

{We consider the base $2 \times 2$ MM below, and consider two Strassen-like algorithms due to Strassen himself and Winorad \cite{WINOGRAD}. The submatrix multiplications used in the Strassen algorithm are shown by $S$s and the submatrix multiplications used in the Winograd algorithm are shown by $W$s. The base $2 \times 2$ MM  $\textbf{C} = \textbf{A}^T\textbf{B}$ is given by 
\[
\underbrace{
\begin{bmatrix}
    C_{11}       & C_{12} \\
    C_{21}       & C_{22} 
\end{bmatrix}
}_{\mathbf{C}} =
\underbrace{
\begin{bmatrix}
    A_{11}       & A_{12} \\
    A_{21}       & A_{22} 
\end{bmatrix}
}_{\mathbf{A^T}} 
\underbrace{
\begin{bmatrix}
    B_{11}       & B_{12} \\
    B_{21}       & B_{22} 
\end{bmatrix}
}_{\mathbf{B}} 
\]

\noindent where sub-matrix computations are given by the following set of equations, 

\begin{center}
\begin{tabular}[h!]{l@{} l@{}}
 \(\mkern-24mu S_1  \! \! =  \! \! (A_{11} \!\! + \!\! A_{22}) \! (B_{11} \!\! + \!\! B_{22})\)&   \( \quad W_1 \!\! = \!\! A_{11}B_{11}\) \\
 \(\mkern-24mu S_2  \! \! =  \! \! (A_{21} \!\! + \!\! A_{22}) \!  B_{11}\) 				   &   \( \quad W_2 \!\! = \!\! A_{12}B_{21}\) \\
 \(\mkern-24mu S_3  \! \! =  \! \! (B_{12} \!\! - \!\! B_{22}) \!  A_{11}\) 				   &   \( \quad W_3 \!\! = \!\! A_{22}\!(B_{11} \!\!-\!\! B_{12} \!\!-\!\! B_{21} \!\!+\!\! B_{22}) \)\\
 \(\mkern-24mu S_4  \! \! =  \! \! (B_{21} \!\! - \!\! B_{11}) \!  A_{22}\) 				   &   \( \quad W_4 \!\! = \!\! (A_{11} \!\! - \!\! A_{21}) \! (B_{22} \!\! - \!\! B_{12}) \)\\
 \(\mkern-24mu S_5  \! \! =  \! \! (A_{11} \!\! + \!\! A_{12}) \!  B_{22}\) 				   &   \( \quad W_5 \!\! = \!\! (A_{21} \!\! + \!\! A_{22}) \! (B_{12} \!\! - \!\! B_{11}) \)\\
 \(\mkern-24mu S_6  \! \! =  \! \! (A_{21} \!\! - \!\! A_{11}) \! (B_{11} \!\! + \!\! B_{12})\)&   \( \quad W_6 \!\! = \!\! B_{22} \! (A_{11} \!\! + \!\! A_{12} \!\! - \!\! A_{21} \!\! - \!\! A_{22}) \)\\
 \(\mkern-24mu S_7  \! \! =  \! \! (A_{12} \!\! - \!\! A_{22}) \! (B_{21}\!\!+ \!\! B_{22}) \) &   \( \quad W_7 \!\! = \!\! (A_{11} \!\! - \!\! A_{21} \!\! - \!\! A_{22})\!(B_{11} \!\!- \!\! B_{12} \!\!+\! \! B_{22})\)
\end{tabular}
\end{center}

\noindent from which we can compute sub-matrices of the multiplication outputs, namely $C_{11},C_{12},C_{21}, C_{22}$ as given by
\begin{eqnarray}
\mkern-15mu C_{11} &=& S_1+S_4-S_5+S_7=W_1+W_2 \label{eqn1}\\
\mkern-15mu C_{12} &=& S_3+S_5=W_1+W_5+W_6-W_7 \label{eqn2}\\
\mkern-15mu C_{21} &=& S_2+S_4=W_1-W_3+W_4-W_7 
\label{eqn3}\\
\mkern-15mu C_{22} &=& S_1-S_2+S_3+S_6=W_1+W_4+W_5-W_7
\label{eqn4}
\end{eqnarray}
\hspace{5mm}If only Strassen or Winograd algorithms were used to calculate $\textbf{C}$ matrix above with parallel setting it would require 7
independent compute nodes. In order to compute all necessary sub-matrices we shall need all computations completed on time. However due to potential stragglers, the final merge (execution of equations \eqref{eqn1}, \eqref{eqn2}, \eqref{eqn3} and \eqref{eqn4}) may be delayed. To overcome this difficulty, a trivial approach is replication (2-copy), which would require 14 compute nodes. However, although average performance will increase using this approach, compute nodes which are calculating the same sub-matrix multiplication may be subject to the presence of stragglers, which would still result in a delayed operation. 

\begin{figure}[t!]
\includegraphics[width=12cm, height=7cm]{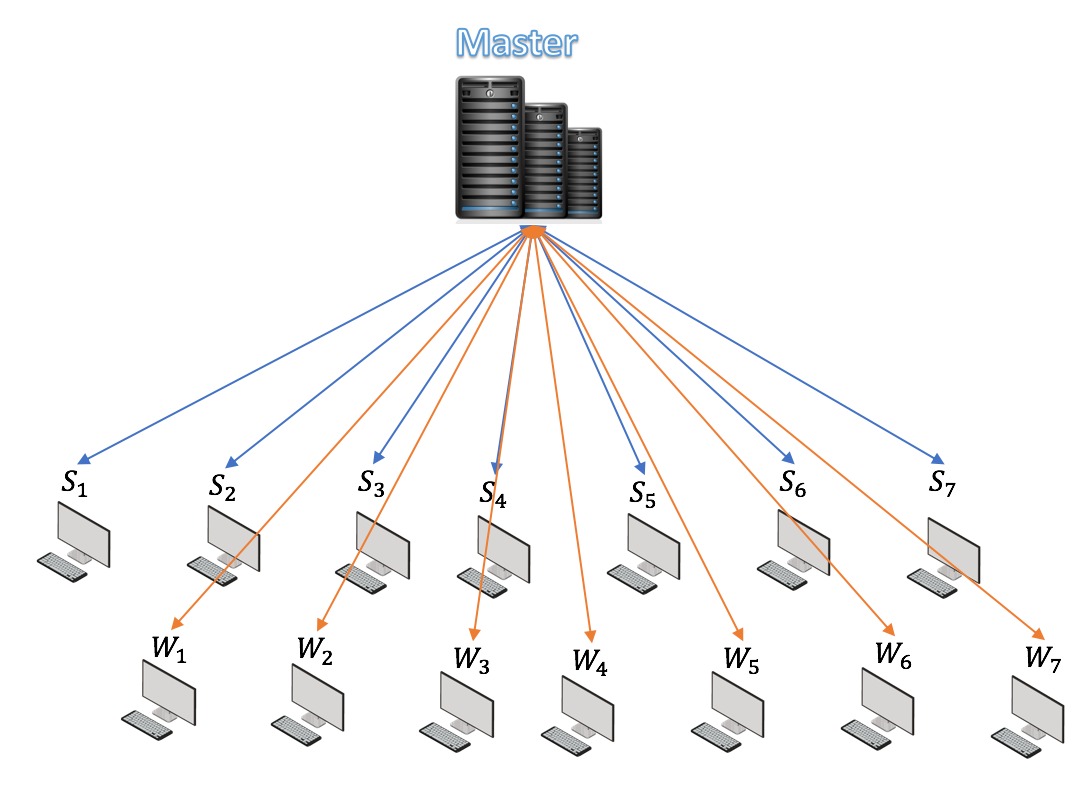}
\centering
 \small{\caption{Master-Slave Configuration}}
\label{figure:fig1}
\end{figure}

\subsection{Proposed Method}
Our proposed technique is based on using two distinct Strassen-like algorithms instead of replicating the same algorithm.  This would also require 14 compute nodes. This approach on the other hand can be shown to generate   \textit{local computations} that relate some of the sub computations of the distinct algorithms which will help with the delay performance. For instance, suppose $S_2, S_5, W_2$ and $W_5$ are all the delayed multiplications. In the pure replication method, this case cannot be recovered quickly. However with the proposed method, first $S_2$ is recovered using equation (3). Then, using equation (4), $W_5$ can be recovered using the result of $S_2$ from the first recovery phase. Next, $S_5$ is recovered using equation (2) and finally, equation (1) recovers $W_2$ and eventually the actual MM output can be reconstructed without waiting for the delayed computations. 

In our study, we consider a master-slave configuration as shown in Fig. 1. The master node is responsible for encoding and distributing multiplications to compute nodes and decoding the multiplication results from all available worker nodes, whereas worker nodes (slaves) are responsible for executing the given multiplication. In addition to equations \eqref{eqn1}, \eqref{eqn2}, \eqref{eqn3} and \eqref{eqn4}, sub-matrix computations of $C$ can be found from different combinations of various compute node outputs, some which are given by
\begin{eqnarray}
\mkern-15mu C_{11} &=& S_2+S_4-S_6+S_7+W_4-W_6  \\
\mkern-15mu C_{12} &=& S_1+S_3+S_4+S_7-W_1-W_2  \\
\mkern-15mu C_{21} &=& S_2+S_3+S_4+S_5-W_1-W_5-W_6+W_7  \\
\mkern-15mu C_{22} &=& S_3+S_5+W_4-W_6 
\end{eqnarray}
\hspace{5mm}These new relations lead to more \textit{local computation} possibilities, which may be used to help some of the delayed multiplications as shown in our example. Through computer-aided search, we have generated all local computation possibilities. Unfortunately, these local computations are not enough to recover all delayed sub-matrix multiplications (only a subset). Therefore, in order to help with the overall delay issue,  parity (extra) sub-matrix multiplications (PSMMs) can be generated at the expense of increasing communication/computation cost of the system. To jointly find all possible local computations involving the sub-marices of $C$ and PSMMs, we have written a computer procedure. Before providing the algorithm details, let us introduce some notations.

\begin{algorithm}[t!]
\textbf{Input:} $M:$ Number of Sub-Matrix Multiplications\\
\hspace*{6.3ex} $K:$ Number of Combinations\\
\textbf{Initialize:}
$C_{11}:$ $0$x$8040$, 
$C_{12}:$ $0$x$0804$,
$C_{21}:$ $0$x$2010$, 
$C_{22}:$ $0$x$0201$\\
\textbf{Output:} $L:$ Local Computations \\
\hspace*{6.3ex}  \, $P:$ Parity Calculations\\
\SetKwFunction{FMain}{SearchLP}
\SetKwProg{Fn}{Function}{:}{}
    \Fn{\FMain{$M, K$}}
    { 
        $N \leftarrow \binom{M}{K}$ \\ 
        $AllComb \leftarrow all \ combinations$ \\
        \For{$n = 1:N$}{
            $Comb \leftarrow AllComb (n,:)  $ \\
            \For{$m=0:2^{K}-1$}
            {
            $(n_1...n_K)_2 = (m)_{10} $ \\
            $Comb \leftarrow Comb \odot ((-1)^{n_1}...(-1)^{n_K}) $ \\
            \eIf{$Comb= C_{11} || C_{12} || C_{21} || C_{22} $}{
                 $L \leftarrow L+Comb  $ 
                
            }
             {
                  \If{$Comb = one \ multiplication$}{
                 $P \leftarrow P+Comb  $ 
                
            }
             }
            
          }  
        }
    }

\textbf{return} $L, P$
\caption {Algorithm for Finding All Local Computations and Local Parity (LP) Calculations}
\label{algo:1}
\end{algorithm}

Firstly, sub-matrix multiplications of $A$ and $B$ matrices can be represented in a matrix form  as shown at Table I. Then, sub-matrices of $C$ can be written in terms of $4 \times 4$ binary matrices in which a logical one would indicate the presence of the corresponding term in Table I. We vectorize (column-wise, MSB on top) these matrices and express them in hexadecimal forms. For example $C_{11}=A_{11}B_{11} + A_{21}B_{12} = [\boldsymbol{e}_1^T \ \boldsymbol{0}^T  \ \boldsymbol{e}_2^T  \ \boldsymbol{0}^T ] = 0x8040$ where $\boldsymbol{e}_i$ is the all-zero vector $\boldsymbol{0}$ except the $i$-th entry is 1. With this short-hand notation, our algorithm can be given as described in Algorithm \ref{algo:1} where $\odot$ stands for the Hadamard product. 


In this study we employ Strassen's and Winograd's original schemes as one of the two Strassen-like algorithms, our proposed method is applicable for any-number Strassen-like algorithms which can be efficiently  searched  using \textit{Triple Product Condition}. Due to space limitation, we refer the reader to \cite{Karstadt} for further discussions.
\begin{table}[t]
\normalsize
\centering
\begin{tabular}{l||l|l|l|l|}
\cline{2-5}
 & $A_{11}$ & $A_{12}$ & $A_{21}$  & $A_{22}$  \\ \hline \hline
\multicolumn{1}{|l|}{$B_{11}$} & $A_{11}B_{11}$ & $A_{12}B_{11}$ & $A_{21}B_{11}$ & $A_{22}B_{11}$  \\ \hline
\multicolumn{1}{|l|}{$B_{12}$} & $A_{11}B_{12}$ & $A_{12}B_{12}$ & $A_{21}B_{12}$ & $A_{22}B_{12}$  \\ \hline
\multicolumn{1}{|l|}{$B_{21}$} & $A_{11}B_{21}$ & $A_{12}B_{21}$ & $A_{21}B_{21}$ & $A_{22}B_{21}$  \\ \hline
\multicolumn{1}{|l|}{$B_{22}$} & $A_{11}B_{22}$ & $A_{12}B_{22}$ & $A_{21}B_{22}$ & $A_{22}B_{22}$  \\ \hline
\end{tabular}
\caption{Multiplication of $A$ and $B$ elements}
\end{table}

\begin{table}[t!]
\centering
\begin{tabular}{||c||}
\hline
 $C_{11}$ \\
 \hline \hline
 $S_2+S_4+W_2+W_3-W_4+W_7$ \\
 $S_3+S_5+W_2-W_5-W_6+W_7$ \\
 $S_1-S_2+S_3+S_6+W_2-W_4-W_5+W_7$ \\
 $S_1-S_2+S_3+S_7-W_3+W_4-W_5-W_6$ \\
 $S_1-S_2-S_5+S_6+W_1+W_2-W_4+W_6$ \\
 $S_1-S_2-S_5+S_7+W_1-W_3+W_4-W_7$ \\
 $S_1+S_3+S_4+S_7-W_1-W_5-W_6+W_7$ \\
 $S_2-S_3+S_4-S_5-S_6+S_7+W_1+W_4+W_5-W_7$ \\
 $S_2-S_3+S_4-S_5+W_1+W_2+W_3-W_4+W_5+W_6$ \\
 \hline
\end{tabular}
\caption{Additional local relations involving $C_{11}$.}
\end{table}
\section{Numerical Results and Discussions}

\begin{figure}[t!]
\includegraphics[width=13cm, height=10cm]{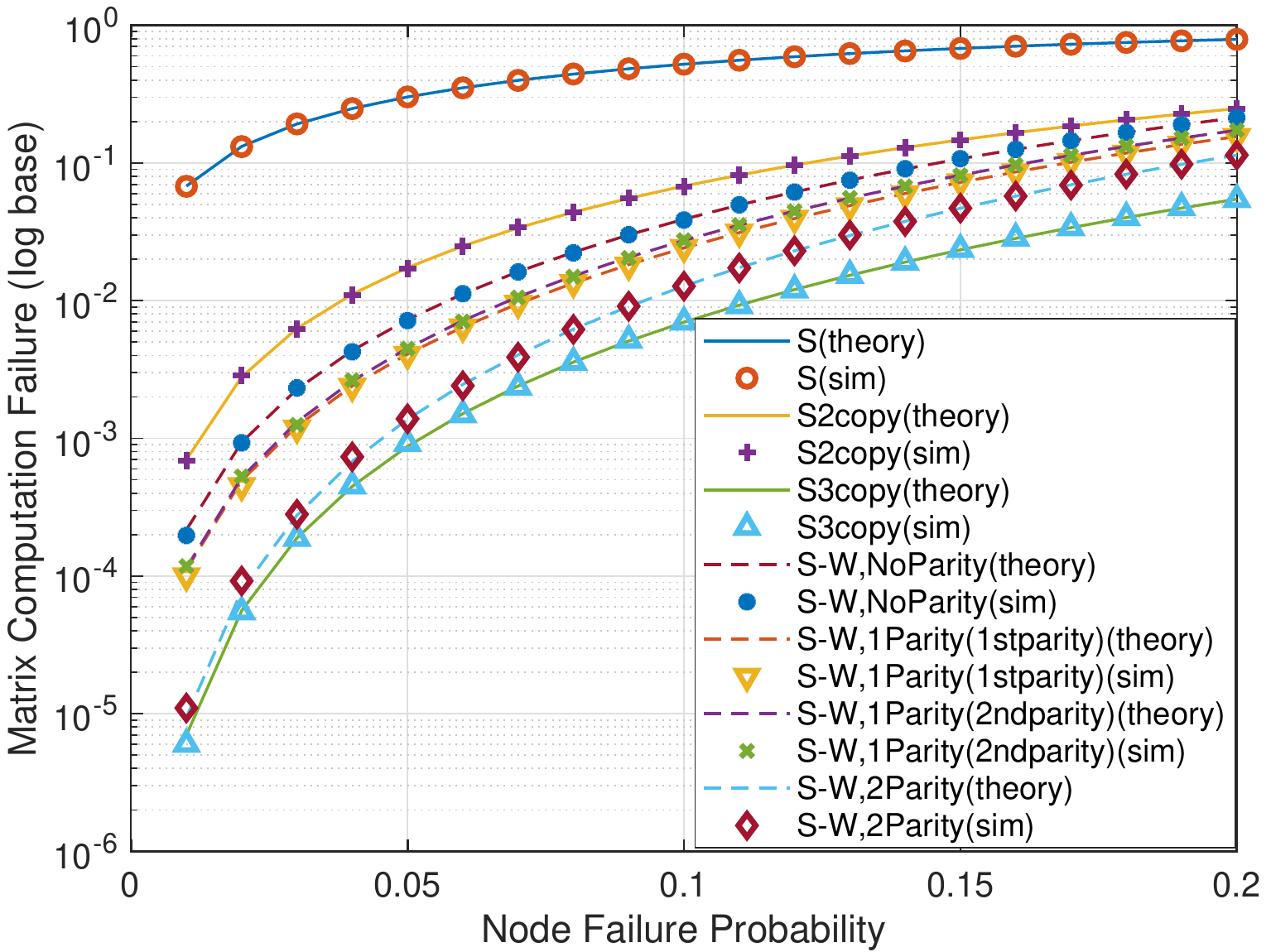}
\centering
 \footnotesize{\caption{Comparison of different methods as a function of node (sub-computation) failure probability, S: Strassen, W: Winograd.}}
 \label{fig22:perf}
\end{figure}

In this section, numerical comparison of the probability of computation completion of simple replication and the proposed methods are compared for two specific base $2 \times 2$ Strassen-like algorithms. For simplicity, we assume compute nodes either complete their assigned task on time or fail to do so independent of each other. More sophisticated methods such as exponential work completion time to model the delayed computations are left for future studies. Monte Carlo simulations as well as theoretical calculations for the simple failure model are also provided.

Our simulation results are provided in Fig. 2, where proposed schemes either use no, one or two PSMMs. In the case of no parity computations, we combine Strassen's and Winograd's subcomputations (a total of 14) to search for local relations between them. We were able to find 52 independent relations some of which are presented in equatiıons (1)-(8). For instance in addition to equations (1) and (5), more local relations for $C_{11}$ are listed in Table II.  Among the parity sub-computations, $2^{nd}$ PSMM  corresponds to the identical copy of $W_2$, and $1^{st}$ PSMM corresponds to $S_3+W_4 = A_{21}(B_{12}-B_{22})$ which is found using our computer-aided search routine. We note that these choices are not arbitrary. If no PSMM is used and simultaneous local computations of the pairs ($S_3$, $W_5$) or ($S_7$, $W_2$) are delayed, fault tolerant computation is not sufficiently achieved. To be able to recover at least one of the delayed multiplications in the first pair, a PSMM which involves the delayed subcomputation needs to be found. This turns out to be the $1^{st}$ PSMM presented above. On the other hand for the second pair, there is no PSMM which involves just $S_7$ or $W_2$. So for the $2^{nd}$ PSMM, $S_7$ or $W_2$ are equally well redundant subcomputations. We have arbitrarily have chosen $W_2$ as our  $2^{nd}$ PSMM.


In Fig. 2, theoretical results are also presented based on the following equations.  Let $M$ be number of compute nodes, $p_e$ to be the node failure probability and $FC(k)$ to be $k$-failure (delayed) combinations of nodes such that $C$ matrix cannot be completely recovered. Due to independence of node failures, the probability of reconstruction failure can be expressed in general as
\begin{equation}
    P_f = \sum_{k=1}^{M}FC(k) p_e^k  (1-p_e)^{M-k}.
\end{equation}
For the replication methods (e.g. Strassen with a single, double and tripple copies), which do not bear any parity computations, $FC(k)$ can be expressed in a closed form as shown below
\begin{align}
    FC(k) &= \sum_{n=1}^{\left \lfloor{\frac{k}{c}}\right \rfloor} (-1)^{n+1}\binom{7}{n}\binom{7c-cn}{k-cn}\mathbf{1}_{(k\geq c)} 
\end{align} 
where $c$ is the number of copies (redundant computations), $\mathbf{1}_{(k\geq c)}$ is the indicator function that equals to 1 if $k \geq c$, 0 otherwise. Both expressions can be proved by induction. Also, note that $FC(k)$ in equation (11) for single copy can be reduced to $\binom{M}{k}$ for  $k=1,2,...M$ where $M=7$ due to the computation output of any node being delayed would result in a reconstruction failure of the matrix $C$. Finally, for the proposed schemes with parity computations, finding a close form expression for $FC(k)$ depends on the local relations we found and turns out to be hard to express in a closed form. Therefore, $FC(k)$'s are calculated with the aid of a computer.

Numerical results indicate that the proposed scheme (in particular Strassen-Winograd pair) either outperforms replication schemes or shows close performance using less redundant computations. For instance, our proposed  method (with 2 PSMMs), performs very close to three-copy Strassen and yet requires less compute nodes i.e., $2 \times 7 + 2 = 16$ compute nodes compared to $3 \times 7 = 21$, i.e., providing $24\%$ reduction in computation abilities of the system. Also, effectiveness of adding PSMMs (at the expense of increased communication cost) can be clearly observed from the performance difference between the proposed schemes with/out PSMMs.

\section{Conclusions and Future Work}

Although our proposed scheme is general and applicable to any pair of Strassen-like algorithms, we only have shown results using Strassen's and Winograd's original algorithms. However, this does not mean we obtain the best results with this selection as better Strassen-like pairs that can generate more independent local relations may be found  using the Triple Product Condition mentioned earlier. Finally, we have to note that our scheme is based on computer-aided efficient search procedures for local relation enumerations. However, we are also actively looking for systematic redundancy generation schemes for recursive matrix multiplication methods as well as deep learning techniques recently fueled by DeepMind initiatives for faster matrix multiplications \cite{fawzi2022discovering}.

\bibliographystyle{ieeetr}
\bibliography{references}

\end{document}